\documentstyle[aps,prl,twocolumn]{revtex}
\newcommand{\bea}{\begin{equation}}
\newcommand{\eea}{\end{equation}}
\newcommand{\ber}{\begin{eqnarray}}
\newcommand{\eer}{\end{eqnarray}}
\begin{document}
\draft
\wideabs{
\title{\bf Non-equilibrium growth in a restricted-curvature model}
\author{Amit Kr. Chattopadhyay}
\address{
Dept. of Theoretical Physics, \\
Indian Association for the Cultivation of Science, \\
Jadavpur, Calcutta 700 032}
\maketitle
\date{}
\begin{abstract}
The static and dynamic roughenings of a growing crystalline facet is studied
where the growth mechanism is controlled by a restricted-curvature (RC)
geometry. A continuum equation, in analogy with the Kardar-Parisi-Zhang
(KPZ) equation is considered for the purpose. It is shown here that although
the growth process begins with a RC geometry, a structural phase transition
occurs from the restricted-curvature phase to the KPZ phase. An estimation
of the corresponding critical temperature is given here. Calculations on the
static phase transition give results along the same line as existing predictions,
apart from minor numerical adjustments. \\\\
PACS number(s): 05.40.+j, 68.35.Rh, 64.40.Ht
\end{abstract}
}
Growth of interfaces in a strongly temperature controlled regime has been the
subject of a considerable portion of recent technological developments [1].
The interest in this field has mainly been generated by the microscopic
roughness that originates as a result of competition among different effects,
such as surface tension, thermal diffusion and different noise factors coming
into play during the growth process. Even in equilibrium, a crystalline facet
"remains practically flat until a transition temperature is reached, at which
the roughness of the surface increases very rapidly" [2]. Numerous theoretical
models, starting from Burton, etal [2] have been proposed to account for a
detailed analysis of the height fluctuations during the growth process on
both sides of the roughening temperature $ T_R $ [3-5]. They found a change
in mobility from activated growth in the nucleated phase (characterized by
low temperatures) to nonactivated growth at large temperatures. The
pinning-depinning growth model of Chui and Weeks [3] has later been
numerically extended to a polynuclear growth model by Sarloos and Gilmer [6]
confirming a growth by island formation which is continuously destroyed by
any small chemical potential favoring the growth. Theoretical forays in this
front have continued with detailed predictions on the equilibrium roughening
transition [7,8] which have also found experimental justifications in [9].
\par
Starting from the early theoretical efforts [2,3] to the present day
developments [1,10], numerous discrete [11,12] and continuum [13,14] models
have been proposed to study both equilibrium and nonequilibrium surface
properties. However all
these different models seem to fall within either of the KPZ or the Lai-Das
Sarma universality class, barring a few exceptions [11,15]. The atomistic
growth process of the latter type deals with surfaces grown by molecular beam
epitaxy (MBE) method, whose distinguishing feature is that growth occurs
under surface diffusion conditions, with the deposited atoms relaxing to the
nearby kinks. The essential idea employed was to modify the relaxation
mechanism as a locally surface minimizing curvature, instead of surface area.
To linear order, this mechanism was supposed to mimic the growth dynamics of
crystalline surfaces [12,14].
\par
Alternative efforts in this front have mainly centered around the development
of theoretical models whose dynamics can be mapped to either of these two main
universality classes. A classic example is an equilibrium restricted-
curvature (RC) model studied by Kim and Das Sarma [16]. The corresponding
growth rule describes a growth restricted on the local curvature,
$ \mid{{\nabla}^2 h}\mid \leq N $, and is obeyed at both the growing site
and its nearest neighbors where $ N $ is any fixed positive integer. We adopt
the nonequilibrium class of growth proposed by Kim and Das Sarma as the
starting point of our study of the dynamics of the nonequilibrium growth in a
MBE process.
\par
Considering two dimensional growth pertaining to a lattice structure where
$ h(\vec r,t) $ is the height of the interface at time $ t $ at position
$ \vec r $, the equation goes like

\ber
\eta \frac{\partial h}{\partial t} &=& -\gamma {\nabla}^4 h(\vec r,t) +
\frac{\lambda}{2} {\mid {\nabla}^2 h(\vec r,t) \mid}^2  \nonumber \\
& & -\frac{2\pi V}{a}
sin[\frac{2\pi} h(\vec r,t)] + F + R(\vec r,t)
\eer

where $ \eta $ is the inverse mobility which fixes the time scale, $ \gamma $
is the surface tension, $ a $ is the lattice constant and $ V $ is the
strength of the pinning potential. $ F $ is a steady driving force with $ R $
the white noise defined as

\bea
< R(\vec r,t) R(\vec r',t') > = 2D {\delta}^2(\vec r-\vec r') \delta(t-t')
\eea

The biharmonic term on the right side of the above equation gives the basic
relaxation mechanism in operation. The second term is the most important term
for the restricted-curvature (RC) dynamics and has been incorporated in
analogy with the KPZ equation [13]. Just as the lateral term $ {\mid{\vec
\nabla h}\mid}^2 $ gives the nonlinearity in a KPZ growth, the curvature
dependence in our model is proposed to produce a leading order nonlinearity
of the form $ ({\nabla}^2 h)^2 $. From phenomenological considerations,
since the discretisation  scheme prohibits a high curvature, the constant
$ \lambda $ is negative. The third term comes along due to the lattice
structure preferring integral multiples of $ h $, which means that $ h $ is
measured in units of $ a $. $ F $ is the steady driving force required to
depin a pinned interface while all the microscopic fluctuations in the growth
process are assimilated in the noise term $ R $. It can be shown easily that
the second and third terms are obtained from a variant of the sine-Gordon
Hamiltonian

\bea
E[h(\vec r,t)] = \int\int d^2r [\frac{\gamma}{2} ({\nabla}^2 h)^2 -
V cos(\frac{2\pi}{a} h)]
\eea

\par
We now start pursuing the most important goal regarding the dynamics of the
growing interface - whether a non-equilibrium phase transition exists or not.
In other words, whether we can specify a critical temperature $ T_R $ below
which the surface is flat and above which the surface starts showing a
dynamic roughening. This obviously brings into question the interplay of
strengths between the non-linear term $ ({\nabla}^2 h)^2 $ and the sine-
Gordon potential. Three limits of the above eqn.(1) are well known: \\
(i) $ \lambda = V = F = 0 $; characterizing a stationery interface [11,12,16].
This sort of growth has found experimental justification in the works of Yang,
etal and Jeffiies, etal [19]. \\
(ii) $ \lambda = F = 0,\:\:V \neq 0 $; characterizing the growth dynamics of
crystalline tensionless surfaces [20]. This model in equilibrium depicts a
roughening transition to the high temperature regime of the sine-Gordon model
and is expected to modelise the vacuum vapor deposition dynamics by MBE growth
[21]. \\
(iii) $ V = F = 0,\:\:\lambda \neq 0 $; a very special case of the numerical
simulation predicting a "local model" for dendritic growth with analytics
failing to define a stable, non-trivial fixed point [22]. This issue is still
under much debate and really needs a deeper understanding to have any final
say in the matter.
\par
However in both the first and second situations, detailed numerical analysis
have predicted a transition from the conserved MBE growth to the Edwards-
Wilkinson [23] type phase characterized by the generation of a
$ {\nabla}^2 h $ term [20,24]. The generation of this so-called "surface
tension" term in the dynamics although surely being a fall-out of the
renormalization of the sine-Gordon potential in the latter model, actually
demands a greater attention. Combining this idea with the proposition put
forward by Kim and Das Sarma [16], we therefore pose the most general
situation concerning a surface growing under MBE and try to ascertain the
associated roughening process throughout the whole temperature range, with
the growth essentially occuring under a surface curvature constraint. In the
following analysis, we employ standard dynamic renormalization techniques to
probe the dynamics of our proposed Langevin equation (1) [7,8] and later on
follow the line of Chui and Weeks [3,7] in taking account of the static renormalization of the model Hamiltonian shown in eqn.(3).
\par
As usual, we start by first integrating over the momentum shell
$ \Lambda(1-dl) < \mid \vec k \mid < \Lambda $ perturbatively in $ \lambda $
and $ V $. Thereafter we rescale back the variables in the form
$ \vec k \rightarrow \vec k' = (1+dl)\vec k,\:\:h \rightarrow h' = h $ and
$ t \rightarrow t'=(1-4dl)t $. The various coefficients follow the rescaling
$ \eta \rightarrow \eta' = \eta,\:\:\gamma \rightarrow \gamma' = \gamma,
\:\:\lambda \rightarrow \lambda' = \lambda,\:\:V \rightarrow V' = (1+4dl)V,
\:\:F \rightarrow F' = (1+4dl)F $. We set up the perturbative scheme to
rewrite eqn.(2) in a comoving frame moving with velocity $ F/\eta $ as

\bea
\eta \frac{\partial}{\partial t} h = -\gamma {\nabla}^4 h + \Phi(h) + R
\eea

where $ \Phi(h) = -\frac{2\pi V}{a}\:sin[\frac{2\pi}{a}(h+\frac{F}{\eta} t)]
+ \frac{\lambda}{2} ({\nabla}^2 h)^2 $.
Thereafter going exactly by the analysis of Nozieres and Gallet and Rost and
Spohn [7,8] and including the corrections in [25], we finally arrive at an
expression for the renormalized mode coupling term,

\ber
\Phi^{SG} &=& -\frac{2{\pi}^3 V^2 T}{{\gamma}^2 a^5} dl\:\int_{-\infty}^t\:
\int d^2r'\:\frac{1}{t-t'}\: J_0(\Lambda(\mid \vec r-\vec r'\mid) \nonumber \\ 
& & G_0(\mid \vec r-\vec r' \mid, t-t') 
\times e^{-[\frac{\gamma}{\eta}{\Lambda}^4(t-t') + \frac{2\pi T}{a^2
\gamma} \phi(\mid \vec r-\vec r' \mid,t-t')]} \nonumber \\
& & \times [\frac{2\pi}{a} 
[\frac{\partial}{\partial t} \bar h(\vec r,t) (t-t')
- \frac{1}{2} {\partial}_i {\partial}_j \bar h(\vec r,t) (r_i-r_i') \times
\nonumber \\
& & (r_j-r_j')]\:cos[\frac{2\pi}{a} \frac{F}{\eta} (t-t')] 
+ [1-\frac{2{\pi}^2}{a^2} [{\partial}_i \bar h(\vec r,t)]^2 \nonumber \\
& & \times (r_i-r_i')^2]\:sin [\frac{2\pi}{a} \frac{F}{\eta} (t-t')]
\eer

where $ \Lambda \sim 1/a $ is a suitably chosen upper cut-off with the
Green's function $ G(x,t) $ given by

\bea
G(x,t)= \int dk\:e^{ikx-\nu t k^4} 
\eea

and 

\bea
\phi(\tilde \rho,x) = \int_0^{\Lambda}\:\frac{dk}{k}\:[1-J_0(k{\tilde \rho})]\:
e^{-\frac{\gamma}{\eta} k^2 (t-t')}
\eea

with $ \tilde \rho = \Lambda \rho $ and $ x = \frac{\gamma(t-t')}
{\eta {\rho}^2} $.
\par
Turning now to the above eqn.(5), we find that starting with a structurally
non-linear term in the Langevin equation, the dynamic renormalization has
initiated a mode coupling structure with quite a different composition.
The absence of the nonlinear $ ({\nabla}^2 h)^2 $ term in $ {\Phi}^{SG} $
implies that once starting with a restricted curvature model of growth, The
lattice structure partakes a dynamics where in a finite time after the start,
$ \lambda $ is renormalized to zero and thereafter the KPZ type [13]
nonlinearity takes over. Thus a competition ensues between the alternate
pinning and depinning forces offered by the $ (\nabla h)^2 $ and
$ ({\nabla}^2 h)^2 $ nonlinearities. Also the production of the surface
tension term $ {\nabla}^2 h $ ensures a dynamic phase transition from the
restricted-curvature regime to the KPZ regime. From an analysis of the
following renormalization flows, we arrive at an expression for the
temperature at which the transition occurs:

\bea
\frac{dU}{dl} = (4-n)U
\eea

\bea
\frac{d\gamma'}{dl} = \frac{2{\pi}^4}{\gamma a^4}\:n A^{(\gamma')}
(n;\kappa)\:U^2
\eea

\bea
\frac{d\gamma}{dl} = \frac{{\pi}^4}{6\gamma a^4}\:n A^{(\gamma)}(n;\kappa)\:
U^2
\eea

\bea
\frac{d\eta}{dl} = \frac{8{\pi}^4}{\gamma a^4}\:\frac{\eta}{\gamma}\:n
A^{(\eta)}(n;\kappa)\:U^2
\eea

\bea
\frac{d\lambda}{dl} = 0
\eea

\bea
\frac{d\lambda'}{dl} = \frac{8{\pi}^5}{\gamma a^5}\:n A^{(\lambda')}
(n;\kappa)\:U^2
\eea

\bea
\frac{dK}{dl} = 4K + \frac{D}{4\pi \eta \gamma}\:\lambda - \frac{4{\pi}^3}
{\gamma a^3}\:n A^{(K)}(n;\kappa)\:U^2
\eea

\bea
\frac{dD}{dl} = \frac{1}{8\pi}\:D\:\frac{{\lambda}^2 D}{{\gamma}^3} +
\frac{8{\pi}^4}{\gamma a^4}\:\frac{D}{\gamma}\:n A^{(\eta)}(n;\kappa)\:U^2
\eea

where $ U = \frac{V}{{\Lambda}^2},\:\:K = \frac{F}{{\Lambda}^2},\:\:
\kappa = \frac{2\pi K}{a \gamma} $ and $ n = \frac{\pi T}{\gamma a^2} $, with
$ \gamma' $ representing the coefficient of the renormalized surface term
[26] and $ \lambda' $ denoting the coefficient corresponding to the KPZ
nonlinearity generated on account of renormalization. $ A^{(i)}(n;\kappa) $
stand as shorthand representation for the integrals, where $ i = \eta,
\lambda', \lambda, \kappa $, etc. and whose detailed functional forms are
given below:

\ber
A^{(\gamma')}(n;\kappa) &=& \int_0^{\infty}\:\frac{dx}{x}\:\int_0^{\infty}\:
d{\tilde \rho}\:{\tilde \rho}^3\:J_0(\tilde \rho) \times cos[\frac{2\pi}{a}
\frac{K x {\tilde \rho}^2}{\gamma}] \nonumber \\
& & \times \int\:dp\:e^{\frac{\tilde \rho}{\lambda}(ip - \frac{\nu}{\Lambda}
\frac{\eta}{\gamma} {\tilde \rho} x p^4)} \times e^{-[x{\tilde \rho}^2 + 2n
\phi(\tilde \rho,x)]}
\eer

\ber
A^{(\eta)}(n;\kappa) &=& \int_0^{\infty}\:dx \int_0^{\infty}\:d{\tilde\rho}\:
{\tilde \rho}\:J_0(\tilde \rho) \times cos[\frac{2\pi}{a} \frac{K x {\tilde
\rho}^2}{\gamma}] \nonumber \\
& & \times \int\:dp\:e^{\frac{\tilde \rho}{\lambda}(ip - \frac{\nu}{\Lambda}
\frac{\eta}{\gamma} {\tilde \rho} x p^4)} \times e^{-[x{\tilde \rho}^2 + 2n
\phi(\tilde \rho,x)]}
\eer
 
\ber
A^{(\lambda')}(n;\kappa) &=& \int_0^{\infty}\:\frac{dx}{x}\:\int_0^{\infty}\:
d{\tilde \rho}\:{\tilde \rho}^3\:J_0(\tilde \rho) \times sin[\frac{2\pi}{a}
\frac{K x {\tilde \rho}^2}{\gamma}] \nonumber \\
& & \times \int\:dp\:e^{\frac{\tilde \rho}{\lambda}(ip - \frac{\nu}{\Lambda}
\frac{\eta}{\gamma} {\tilde \rho} x p^4)} \times e^{-[x{\tilde \rho}^2 + 2n
\phi(\tilde \rho,x)]}
\eer

\ber
A^{(K)}(n;\kappa) &=& \int_0^{\infty}\:\frac{dx}{x}\:\int_0^{\infty}\:
d{\tilde \rho}\:{\tilde \rho}\:J_0(\tilde \rho) \times sin[\frac{2\pi}{a}
\frac{K x {\tilde \rho}^2}{\gamma}] \nonumber \\
& & \times \int\:dp\:e^{\frac{\tilde \rho}{\lambda}(ip - \frac{\nu}{\Lambda}
\frac{\eta}{\gamma} {\tilde \rho} x p^4)} \times e^{-[x{\tilde \rho}^2 + 2n
\phi(\tilde \rho,x)]}
\eer 

\ber
A^{(\gamma)}(n;\kappa) &=& \int_0^{\infty}\:\frac{dx}{x}\:\int_0^{\infty}\:
d{\tilde \rho}\:{\tilde \rho}^5\:J_0(\tilde \rho) \times cos[\frac{2\pi}{a}
\frac{K x {\tilde \rho}^2}{\gamma}] \nonumber \\
& & \times \int\:dp\:e^{\frac{\tilde \rho}{\lambda}(ip - \frac{\nu}{\Lambda}
\frac{\eta}{\gamma} {\tilde \rho} x p^4)} \times e^{-[x{\tilde \rho}^2 + 2n
\phi(\tilde \rho,x)]}
\eer
\par
As already argued, the eqn.(13) giving the flow for the renormalized KPZ
coefficient generates the lattice potential of the driven interface after
the structural phase transition has occured. At some temperature $ T_R =
\frac{4\gamma a^2}{\pi} $,in the units measured, corresponding to the
transition point $ n = 4 $, the roughness dynamics driven by the KPZ force
takes over. From this point the whole dynamics follows in the line predicted
by Nozieres and Gallet [7] and
Rost and Spohn [8], with the biharmonic and structural nonlinear terms
muffled by the surface tension term and the KPZ nonlinearity respectively.
This KPZ regime continues until the growing interface reaches the next
crystalline facet whereon the RC dynamics again comes into play and the whole
mechanism goes on repeating itself throughout the process of nonequilibrium
growth. However in the static case, starting with the standard sine-Gordon
Hamiltonian (eqn.(3)) and again taking clues from [7,8], we arrive at the
following Kosterlitz-Thouless type second order equations [5,10] in terms of
the reduced variables $ y = \frac{4\pi U}{T} $ and $ x = \frac{4\gamma a^2}
{\pi T} $,

\bea
\frac{dx}{dl} = \frac{y^2}{x}\:B(4/x)
\eea

\bea
\frac{dy}{dl} = 4y(1-\frac{1}{x})
\eea

where

\bea
B(n) = \int_0^{\infty}\:d{\tilde \rho}\:{\tilde \rho}^3\:J_0(\tilde \rho)\:
e^{-2n h(\tilde \rho)}
\eea

and

\bea
h(\tilde \rho) = \int_)^{\Lambda}\:\frac{dk}{k^3}\:[1-J_0(k{\tilde \rho}]
\eea

with $ n = \frac{\pi T}{\gamma a^2} $.
The roughening transition occurs at the fixed point $ y = 0,\:x = 1 $ and we
recover all the predictions of Nozieres,  [10] although with slightly
different coefficients. The point to be noted is that although the transition
temperature is twice here compared to that of the standard Kosterlitz-Thouless
(KT) case, both the dynamic and static phase transitions occur at the same
critical temperature $ T_R =\frac{4\gamma a^2}{\pi} $, somewhat alike to the
NG and KT models.
\par
In conclusion, we have studied interface growth models both in the
equilibrium and nonequilibrium cases under a constraint of the growth
occuring under a curvature restriction. The dynamic model gives a phase
transition from the RC growth to the KPZ type at a temperature we have
calculated to be $ \frac{4\gamma a^2}{\pi} $. Experimental data on the
roughening transition of solid ${}^{4}He $ in contact with the superfluid
$ {}^{3}He $ [9] seems to show a crossover to the non-equilibrium  regime in
the superfluid phase. We suspect that in a scenario analogous to the MBE
growth model proposed, the crossover situation is actually mimicked by the
dynamic phase transition found in the model discussed and the larger value of
the critical phase transition temperature reported here might just be the
explanation for the weaker coupling observed around $ 0.1 {}^{o}K $ in
$ {}^{3}He $. However, although the dynamic mechanism proposed here differs
largely from the other existing models, thereby defining a new universality class, the behavior at equilibrium defined by the corresponding static
model is actually a replica of the Kosterlitz-Thouless criticality, the only
variation being in an augmented value of the transition temperature.
\par
The author acknowledges partial financial support from the Council of
Scientific and Industrial Research, India during the whole tenure of this
work. Also sincere gratitude goes to Prof. Jayanta K. Bhattacharjee for
numerous illuminating discussions with him during the course of this work.


\begin{thebibliography}{99}
\bibitem{1} A. L. Barabasi and E. Stanley, "Fractal Concepts in Surface
Growth", Cambridge Univ. Press, NY, 1995.
\bibitem{2} W. K. Burton, N. Cabrera and F. C. Frank, Philos. Trans. R. Soc.
$ \bf 243 A $, 299 (1951).
\bibitem{3} S. T. Chui and J. D. Weeks, Phys. Rev. B $ \bf 14 $, 4978 (1976).
\bibitem{4} H. van Beijeren, Phys. Rev. Letts. $ \bf 38 $, 993 (1977).
\bibitem{5} J. M. Kosterlitz and D. J. Thouless, J. Phys. C $ \bf 6 $, 1181
(1973); J. Phys. C $ \bf 7 $, 1046 (1974).
\bibitem{6} W. van Sarloos and G. H. Gilmer, Phys. Rev. B $ \bf 33 $, 4927
(1986).
\bibitem{7} P. Nozieres and F. Gallet, J. Phys. (Paris) $ \bf 48 $, 353
(1987).
\bibitem{8} M. Rost and H. Spohn, Phys. Rev. E $ \bf 49 $, 3709 (1991).
\bibitem{9} F. Gallet, S. Balibar and E. Rolley, J. Phys. (Paris) $ \bf 48 $,
369 (1987); Y. Carmi, E. Polturak and S. G. Lipson, Phys. Rev. Letts,
$ \bf 62 $, 1364 (1989); P. E. Wolf, F. Gallet, S. Balibar,E. Rolley and
P. Nozieres, J. Phys. (Paris) $ \bf 46 $, 1987 (1985).
\bibitem{10} J. Krug and H. Spohn; P. Nozieres in "Solids Far From
Equilibrium" edtd. by C. Godreche, Cambridge Univ. Press, NY, 1992.
\bibitem{11} S. Das Sarma and P. I. Tamborenea, Phys. Rev. Letts, $ \bf 66 $,
325 (1991).
\bibitem{12} P. E. Wolf and J. Villain, Euro Phys. Letts. $ \bf 13 $, 389
(1990); F. Family and T. Vicsek, J. Phys. A $ \bf 18 $, L75 (1985).
\bibitem{13} M. Kardar, G. Parisi and Y. C. Zhang, Phys. Rev. Letts.
$ \bf 56 $, 889 (1986).
\bibitem{14} Z. W. Lai and S. Das Sarma, Phys. Rev. Letts. $ \bf 66 $, 2348
(1991).
\bibitem{15} J. Krug, Adv. in Phys. $ \bf 46 $, 139 (1997).
\bibitem{16} J. M. Kim and S. Das Sarma, Phys. Rev. E $ \bf 48 $, 2599 (1993).
\bibitem{17} J. Villain, J. Phys. (Paris) I $ \bf 1 $, 19 (1991).
\bibitem{18} S. Das Sarma and S. V. Ghaisas, Phys. Rev. Letts. $ \bf 69 $,
3758 (1992).
\bibitem{19}H. N. Yang, Y. P. Zhao, G. C. Wang and T. M. Lu, Phys. Rev. Letts.
$ \bf 76 $, 3774 (1996); J. II. Jeffries, J. F. Zuo and M. M. Craig, Phys. Rev.
Letts. $ \bf 76 $, 4931 (1996).
\bibitem{20} E. Moro, R. Cuerno and A. Sanchez, Phys. Rev. Letts. $ \bf 78 $,
4982 (1997).
\bibitem{21} A. Zhangwill, J. Cryst. Growth $ \bf 163 $, 8 (1996).
\bibitem{22} N. Govind and H. Guo, J. Phys.  $ \bf 25 $, 5485(1992).
\bibitem{23} S. F. Edwards, D. R. Wilkison, Proc. R. Soc. A $ \bf 381 $, 17
(1982).
\bibitem{24} J. G. Amar, P. M. Lam and F. Family, Phys. Rev. E $ \bf 47 $,
3242 (1993).
\bibitem{25} H. J. F. Knops and L. W. Den Ouden, Physica A $ \bf 103 $, 579 (1980).
\bibitem{26} S. Das Sarma and R. Kotlyar, Phys. Rev. E $ \bf 50 $, R4275 (1994).
\end{thebibliography}
\end{document}